\documentclass[aps,prd,onecolumn,nofootinbib]{revtex4-2} %%,superscriptaddress  %%onecolumn oder twocolumn
\usepackage{titlesec}
\titleformat*{\section}{\LARGE\bfseries}
\titleformat*{\subsection}{\Large\bfseries}
\titleformat*{\subsubsection}{\large\bfseries}%\itshape

\usepackage{epsfig}
\usepackage{enumitem}

\usepackage{graphics}     
\usepackage{graphicx}     
\usepackage{subfigure}
\usepackage{longtable}     
\usepackage{url}          
\usepackage{bm}           
\usepackage{natbib}
\usepackage{textpos}
\usepackage{amsfonts,amsmath,amssymb,mathrsfs,bm,amsthm}
\usepackage{float}
\usepackage{booktabs}
\usepackage{array}
\usepackage{tabu}
\usepackage{dcolumn}
\usepackage{rotating}
\usepackage{ulem}

\newcommand{\RN}[1]{%
  \textup{\uppercase\expandafter{\romannumeral#1}}%
}

\usepackage{nicefrac}
\usepackage{amsthm}
\usepackage{color}
\usepackage{cancel}
\usepackage{appendix}
\usepackage{makecell}
\usepackage{upgreek}

\usepackage[colorlinks=true,linkcolor=blue,citecolor=blue,urlcolor=blue,bookmarksopen]{hyperref}
\newcommand{\appropto}{\mathrel{\vcenter{
  \offinterlineskip\halign{\hfil$##$\cr
    \propto\cr\noalign{\kern2pt}\sim\cr\noalign{\kern-2pt}}}}}
\begin{document}

\title{\Large{
Searching for ultralight scalar dark matter
\\ with muonium and muonic atoms
}}

\author{{\large\bf \rule[30pt]{0pt}{0pt}
Yevgeny~V.~Stadnik 
}}
%\email{yevgenystadnik@gmail.com}

\affiliation{{\large \rule[25pt]{0pt}{0pt}
School of Physics, University of Sydney, Sydney, NSW 2006, Australia
}}

\raggedbottom

%\date{\normalsize{
%\today
%}}

\maketitle

 %% "Manual" Abstract %%
\vspace{3mm}
\begin{center}
\Large{\textbf{Abstract}}
\end{center}
\large

Ultralight scalar dark matter may induce apparent oscillations of the muon mass, which may be directly probed via temporal shifts in the spectra of muonium and muonic atoms. 
Existing datasets and ongoing spectroscopy measurements with muonium are capable of probing scalar-muon interactions that are up to 12 orders of magnitude more stringent than astrophysical bounds. 
Ongoing free-fall experiments with muonium can probe forces associated with the exchange of virtual ultralight scalar bosons between muons and standard-model particles, offering up to 5 orders of magnitude improvement in sensitivity over complementary laboratory and astrophysical bounds.

\vspace{200mm}

\large

%\tableofcontents
%\vspace{200mm}

%%%%%
\textbf{Introduction} --- 
Various astrophysical and cosmological observations at different epochs indicate that the Universe is predominantly composed of dark matter (DM) and dark energy, with only a mere five percent due to the ordinary matter that makes up stars, planets, dust and interstellar gases \cite{PDG_2022_review}. 
Unravelling the identity and microscopic properties of DM remains one of the greatest challenges in modern science. 
Conventional schemes for direct detection of DM have largely focused on possible \textit{particlelike} signatures of weakly-interacting massive particles (WIMPs) with masses in the $\sim \textrm{GeV} - \textrm{TeV}$ range~\cite{Roszkowski_2018_WIMP-review}. 
On the other hand, ultralight bosons with sub-eV masses and a high number density may produce distinctive \textit{wavelike} signatures. 
Ultralight bosons may also resolve other outstanding problems outside of astrophysics; 
for example, the axion (a pseudoscalar particle) may resolve the strong \textit{CP} problem of quantum chromodynamics \cite{Kim-Carosi_2010_Axions-review}, while the relaxion (a scalar particle) may resolve the electroweak hierarchy problem \cite{Graham_2015_Relaxion}. 

There has been growing interest in recent years to search for ultralight scalar DM via its interactions with photons, electrons and nucleons; see \cite{Snowmass_2022_Scalars+Vectors} for a recent overview. 
However, comparatively few studies of the possible interactions of ultralight scalar DM with muons have been undertaken. 
In contrast to electrons and nucleons, there are no stable terrestrial sources of muons. 
As a result, there are no direct stringent constraints on the linear interaction of ultralight scalar bosons with muons from laboratory searches for equivalence-principle-violating forces, which by contrast place very stringent bounds on the linear interaction of ultralight scalar bosons with electrons and nucleons \cite{Rot-Wash_1999_EP,Eot-Wash_2008_EP,Wagner_2012_EP-review,Leefer_2016_EP,MICROSCOPE_2017_EP,Hees_2018_EP,MICROSCOPE_2022_EP}. 
Additional motivation for probing scalar-muon interactions comes from the persistence of various anomalies in muon physics, including the proton radius inferred from measurements of the Lamb shift in hydrogen-like muonic atoms \cite{Pohl_2010_Muonic-H,Antognini_2013_Muonic-H,Pohl_2016_Muonic-D,Krauth_2021_Muonic-He,Pospelov_2011_Proton_radius} and the muon anomalous magnetic moment \cite{BNL_2006_Muon_g-2,FNAL_2021_Muon_g-2,Pospelov_2017_Muon_g-2}. 

In this paper, we point out that spectroscopy measurements with muonium and muonic atoms can provide sensitive novel probes of ultralight scalar DM via its interaction with muons. 
Spectroscopy measurements in muonium \cite{LAMPF_1975_Muonium_HFS-ZF,LAMPF_1999_Muonium_HFS-HF,RAL_2000_Muonium_1S-2S,J-PARC_2021_Muonium_HFS-ZF,PSI_2022_Muonium_2S-2P} and muonic atoms \cite{Pohl_2010_Muonic-H,Antognini_2013_Muonic-H,Pohl_2016_Muonic-D,Krauth_2021_Muonic-He} have provided high-precision tests of bound-state quantum electrodynamics, as well as precise determination of fundamental constants. 
We estimate that existing datasets and ongoing spectroscopy measurements with muonium are capable of probing scalar-muon interactions that are up to 12 orders of magnitude more stringent than astrophysical bounds. 
This opens an exciting new avenue in direct searches for DM. 
We also point out that free-fall experiments with muonium can provide sensitive probes of forces associated with the exchange of virtual ultralight scalar bosons between muons and standard-model particles.

%%%%%
\textbf{Theory} --- 
Ultralight bosons are a good candidate to explain the observed DM. 
Ultralight spinless bosons may be produced non-thermally via the classic ``vacuum misalignment'' mechanism in the early Universe \cite{Wilczek_1983_Axion-cosmo,Sikivie_1983_Axion-cosmo,Fischler_1983_Axion-cosmo} or via the ``thermal vacuum misalignment'' mechanism \cite{Batell_2021_Thermal_misalignment} and can subsequently form a coherently oscillating classical field: 
\begin{equation}
\label{scalar_DM_field}
\phi (t) \approx \phi_0 \cos (\omega_\phi t)  \, , 
\end{equation}
which oscillates at the angular frequency $\omega_\phi \approx m_\phi c^2 / \hbar$ governed approximately by the mass of the DM particle $m_\phi$, with $c$ the speed of light in vacuum, and $\hbar$ the reduced Planck constant. 
Unless explicitly stated otherwise, we adopt the natural system of units $\hbar = c =1$. 
The oscillating DM field in Eq.~(\ref{scalar_DM_field}) carries an energy density, averaged over a period of oscillation, of $\left< \rho_\phi \right> \approx m_\phi^2 \phi_0^2 / 2$. 
In the standard halo model, DM bosons in our local Galactic region are expected to have a root-mean-square speed of $\left< v_\phi^2 \right>^{1/2} \sim 10^{-3}c$ relative to the Galactic Center, with a comparable spread in boson speeds. 
The typical spread in the DM boson energies is hence expected to be $\Delta E_\phi / E_\phi \sim \left< v_\phi^2 \right> / c^2 \sim 10^{-6}$, implying a coherence time of $\tau_\textrm{coh} \sim 2\pi / \Delta E_\phi \sim 10^6 T_\textrm{osc}$, where $T_\textrm{osc} \approx 2\pi / m_\phi$ is the DM period of oscillation. 
In Eq.~(\ref{scalar_DM_field}), we have neglected small motional gradient terms. 
%; i.e., the oscillations of the bosonic DM field are nearly monochromatic, with an associated quality factor of $\sim 10^6$ (relative linewidth of $\sim 10^{-6}$). 
%The field in Eq.~(\ref{scalar_DM_field}) is classical if a large number of DM bosons fit into the reduced de Broglie volume, which for the average local Galactic DM density of $\rho_\textrm{DM} \approx 0.4~\textrm{GeV/cm}^3$ occurs when $m_\phi \lesssim 1~\textrm{eV}$.  {\color{red}Myabe move the reference from here to the start of the section on ``Direct signatures'' later on}

A scalar field $\phi$ can couple to standard-model (SM) fields via the following linear-in-$\phi$ interactions: 
\begin{equation}
\label{linear_scalar_interactions}
\mathcal{L}_\textrm{int}^\textrm{lin} = \frac{\phi}{\Lambda_\gamma} \frac{F_{\mu \nu} F^{\mu \nu}}{4} - \sum_\psi \frac{\phi}{\Lambda_\psi} m_\psi \bar{\psi} \psi  \, , 
\end{equation}
where the first term represents the coupling of $\phi$ to the electromagnetic field tensor $F$, while the second term represents the coupling of $\phi$ to the SM fermion fields $\psi$, with $m_\psi$ the ``standard'' mass of the fermion, and $\bar{\psi} = \psi^\dagger \gamma^0$ the Dirac adjoint. 
The parameters $\Lambda_\gamma$ and $\Lambda_\psi$ denote the respective effective new-physics energy scales. 
The linear couplings in Eq.~(\ref{linear_scalar_interactions}) can be generated, e.g., via the super-renormalisable interaction of $\phi$ with the Higgs field \cite{Piazza_2020_scalar-Higgs} or from \textit{CP}-violating pseudoscalar-fermion couplings \cite{Moody_1984_scalar_CPV,deVries_2022_scalar_CPV}. 
These linear couplings may be absent, however, e.g., as a result of an underlying $Z_2$ symmetry (invariance under the transformation $\phi \to -\phi$). 
In this case, $\phi$ could couple to SM fields via the following quadratic-in-$\phi$ interactions with the effective new-physics energy scales $\Lambda'_\gamma$ and $\Lambda'_\psi$: 
\begin{equation}
\label{quadratic_scalar_interactions}
\mathcal{L}_\textrm{int}^\textrm{quad} = \left( \frac{\phi}{\Lambda'_\gamma} \right)^2 \frac{F_{\mu \nu} F^{\mu \nu}}{4} - \sum_\psi \left( \frac{\phi}{\Lambda'_\psi} \right)^2 m_\psi \bar{\psi} \psi  \, . 
\end{equation}
The interactions in Eqs.~(\ref{linear_scalar_interactions}) and (\ref{quadratic_scalar_interactions}) effectively alter the electromagnetic fine-structure constant $\alpha$ and fermion masses according to \cite{Stadnik_2015_DM-LIFO,Stadnik_2015_DM-VFCs}: 
\begin{equation}
\label{varying_FCs_linear}
\alpha \to \frac{\alpha}{1 - \phi/\Lambda_\gamma} \approx \alpha \left( 1 + \frac{\phi}{\Lambda_\gamma} \right) \, , \,  ~  m_\psi \to m_\psi \left( 1 + \frac{\phi}{\Lambda_\psi} \right)  \, , 
\end{equation}
\begin{equation}
\label{varying_FCs_quadratic}
\alpha \to \frac{\alpha}{1 - (\phi/\Lambda'_\gamma)^2} \approx \alpha \left[ 1 + \left( \frac{\phi}{\Lambda'_\gamma} \right)^2 \right] \, , \,  ~  m_\psi \to m_\psi \left[ 1 + \left( \frac{\phi}{\Lambda'_\psi} \right)^2 \right]  \, , 
\end{equation}
where we have assumed small perturbations. 
In the case of the oscillating scalar field $\phi$ in Eq.~(\ref{scalar_DM_field}), one hence expects the following apparent oscillations in the value of the muon mass $m_\mu$: 
\begin{equation}
\label{oscillating_FCs}
\frac{\Delta m_\mu}{m_\mu} \approx \frac{\phi_0 \cos(m_\phi t)}{\Lambda_\mu} \approx \frac{\sqrt{2 \rho_\phi} \cos(m_\phi t)}{m_\phi \Lambda_\mu} \, , \, \, \frac{\Delta m_\mu}{m_\mu} \approx \frac{\phi_0^2 \cos(2 m_\phi t)}{2 (\Lambda'_\mu)^2} \approx \frac{\rho_\phi \cos(2 m_\phi t)}{m_\phi^2 (\Lambda'_\mu)^2}  \, . 
\end{equation}

In Appendix~\ref{Appendix:A}, we estimate the relative sensitivities of selected transitions in muonium ($e^- \mu^+$ bound state), true muonium ($\mu^- \mu^+$ bound state) and hydrogen-like muonic atoms (bound state of a muon and hadronic nucleus) to variations of $m_\mu$ and other fundamental constants. 
The relative sensitivity coefficients for variations of $m_\mu$ (defined as $\Delta \nu / \nu = K_\mu \Delta m_\mu / m_\mu$) are $|K_\mu| \approx 1$ for most of the considered transitions. 
Notable exceptions arise for the $1S-2S$ interval in muonium ($K_\mu \approx m_e/m_\mu$), the $2S-2P$ Lamb shift in muonium ($K_\mu \approx 3m_e/m_\mu$) and the ground-state hyperfine splitting interval in hydrogen-like muonic atoms ($K_\mu \approx 2$). 

A variety of reference frequencies can be used in measurements of transition frequencies in muonium or muonic atoms. 
For instance, the experiment at RAL \cite{RAL_2000_Muonium_1S-2S,RAL_2000_Thesis-Meyer} used molecular iodine, the ongoing Mu-MASS experiment at PSI \cite{PSI_2018_Muonium_1S-2S} uses GPS-based Cs and Rb microwave atomic clocks \cite{Paolo_2022_Private_Comm}, the experiments at LAMPF \cite{LAMPF_1999_Muonium_HFS-HF,LAMPF_1999_Thesis-Liu} used atomic frequency standards contributing to the LORAN-C navigation system, while the MuSEUM experiment at J-PARC \cite{J-PARC_2021_Muonium_HFS-ZF} used an oven-controlled crystal oscillator \cite{Kanda_2022_Private_Comm}. 
All such conventional reference frequencies are sensitive to variations of $\alpha$ and $m_e$ (and in some cases also to variations of the nucleon and light-quark masses), but they are practically \textit{insensitive} to variations of $m_\mu$. 
Hence in comparisons of a transition frequency in muonium or a muonic atom with a conventional reference frequency, the relative sensitivity of the frequency ratio to variations of $m_\mu$ is simply given by the value of $K_\mu$ for the transition in muonium or muonic atom.

%%%%%
\textbf{Direct signatures} --- 
Let us now appraise the sensitivities of spectroscopy measurements in muonium and muonic atoms to the new-physics energy scales $\Lambda_\mu$ and $\Lambda'_\mu$ appearing in Eqs.~(\ref{linear_scalar_interactions}) and (\ref{quadratic_scalar_interactions}) via the apparent oscillations of $m_\mu$ induced by an ultralight scalar DM field according to Eq.~(\ref{oscillating_FCs}). 
In Appendix~\ref{Appendix:B}, we discuss some general details of searches for signals induced by an oscillating DM field. 
Here we focus on discussion of specific experimental platforms. 
We assume that the interaction of $\phi$ with muons is much stronger than with other SM fields and that $\rho_\phi$ is equal to the average local Galactic DM density, $\rho_\phi = \rho_\textrm{DM} \approx 0.4~\textrm{GeV/cm}^3$ \cite{PDG_2022_review}. 
%Based on Eq.~(\ref{oscillating_FCs}) alone, we can generally expect higher sensitivity to $\Lambda_\mu$ and $\Lambda'_\mu$ at smaller values of $m_\phi$. 

\begin{figure*}[t!]
\centering
\includegraphics[width=8.5cm]{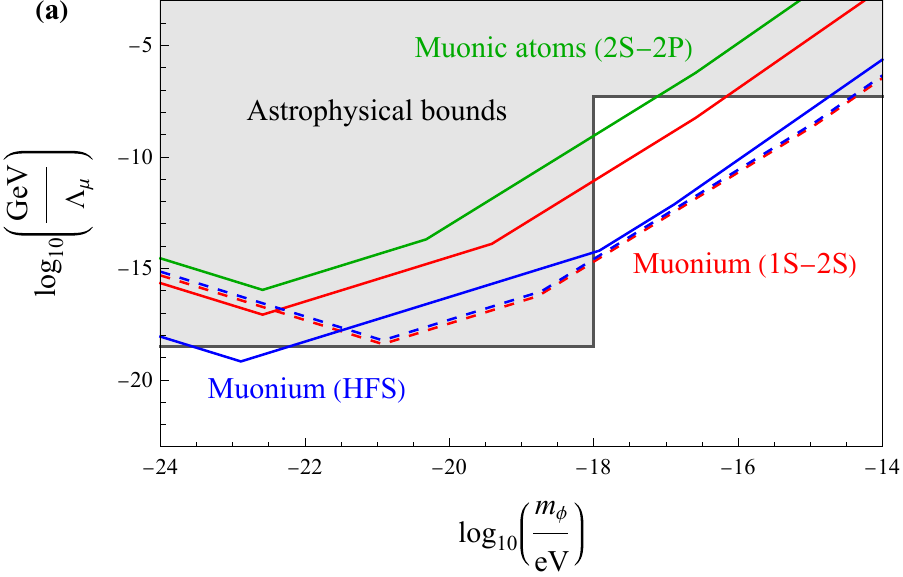}
\includegraphics[width=8.5cm]{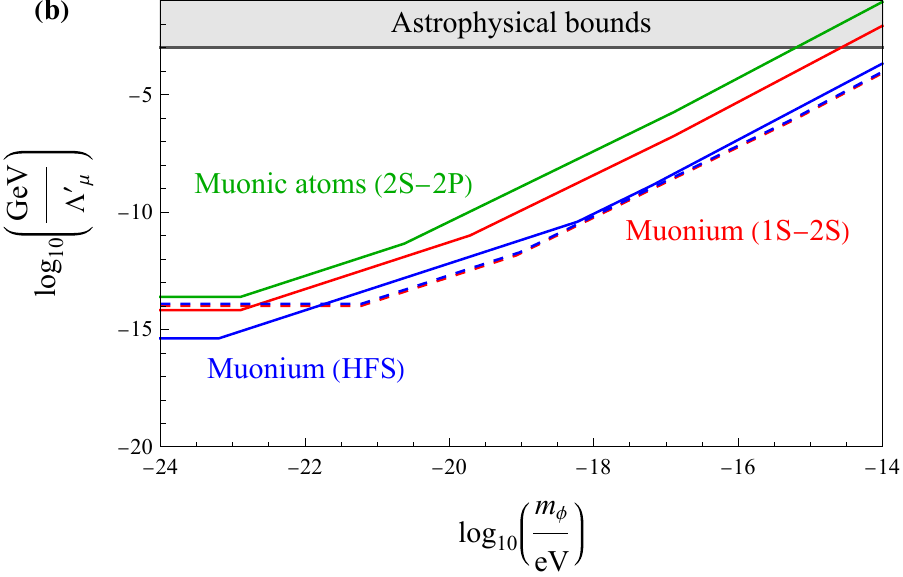}
\caption{ \normalsize (Colour online) 
Estimated sensitivities to \textbf{(a)} the linear scalar-muon interaction parameter $\Lambda_\mu$ in Eq.~(\ref{linear_scalar_interactions}) and \textbf{(b)} quadratic scalar-muon interaction parameter $\Lambda'_\mu$ in Eq.~(\ref{quadratic_scalar_interactions}) of possible searches for apparent oscillations of the muon mass using existing datasets (solid lines) and ongoing measurements (dashed lines) pertaining to spectroscopy measurements of the $1S-2S$ interval in muonium (red), ground-state hyperfine splitting interval in muonium (blue), and $2S-2P$ Lamb shift in muonic atoms (green). 
We assume that $\rho_\phi = \rho_\textrm{DM} \approx 0.4~\textrm{GeV/cm}^3$. 
However, note that for scalar masses $m_\phi \lesssim 10^{-21}~\textrm{eV}$, such scalars cannot account for $100\%$ of the DM \cite{Viel_2017_Fuzzy-DM,Viel_2019_Fuzzy-DM,Marsh_2019_Fuzzy-DM,Schutz_2020_Fuzzy-DM}. 
The sensitivities of the considered spectroscopy experiments to $\Lambda_\mu$ scale as $\propto \sqrt{\rho_\phi}$, meaning that for $\rho_\phi / \rho_\textrm{DM} = \mathcal{O} (0.1)$ corresponding to the maximum allowable fraction when $m_\phi \lesssim 10^{-21}~\textrm{eV}$, the sensitivity weakens by a factor of $\approx 3$. 
Astrophysical bounds are shown by the grey regions. 
Sensitivities and limits are given at the $1\sigma$ level. 
See the main text and Appendices~\ref{Appendix:B} and \ref{Appendix:C} for more details about the sensitivity estimates and complementary limits. 
}
\label{Fig:Scalar_DM_Sensitivity_Plot}
\end{figure*}

%%%
\emph{Muonium:} 
Measurements of the $1S-2S$ interval in muonium were performed at RAL with a precision of $\Delta \nu / \nu \approx 4 \times 10^{-9}$ \cite{RAL_2000_Muonium_1S-2S,RAL_2000_Thesis-Meyer}. 
About $5$ weeks of measurements were performed over a period of $\approx 5$ years, with a single line scan taking $\approx 30-40$ hours and consisting of $\approx 20$ measurements at different frequencies. 
Using Eq.~(\ref{oscillating_FCs}) and noting that $K_\mu \approx m_e/m_\mu$ for the $1S-2S$ transition in muonium, we estimate the sensitivity of this existing dataset to $\Lambda_\mu$ and $\Lambda'_\mu$ as shown by the solid red lines in Fig.~\ref{Fig:Scalar_DM_Sensitivity_Plot}. 
The ongoing Mu-MASS experiment at PSI is targetting an improved precision of $\Delta \nu / \nu \approx 4 \times 10^{-12}$ with $\approx 40$ days of measurements \cite{PSI_2018_Muonium_1S-2S}. 
Assuming that a single line scan takes $\approx 7$ hours \cite{Paolo_2022_Private_Comm}, we estimate the sensitivity of these ongoing measurements to $\Lambda_\mu$ and $\Lambda'_\mu$ as shown by the dashed red lines in Fig.~\ref{Fig:Scalar_DM_Sensitivity_Plot}. 

Low-field measurements of the ground-state hyperfine splitting interval in muonium were performed at LAMPF with a precision of $\Delta \nu / \nu \approx 3 \times 10^{-7}$ \cite{LAMPF_1975_Muonium_HFS-ZF}. 
High-field measurements with an improved precision of $\Delta \nu / \nu \approx 1.2 \times 10^{-8}$ were subsequently performed at LAMPF \cite{LAMPF_1999_Muonium_HFS-HF,LAMPF_1999_Thesis-Liu,LAMFP_1999_Thesis-Reinhard}. 
In the latter case, several months of measurements were performed over a period of nearly a decade, with a single line scan taking $\approx 1$ hour. 
Using Eq.~(\ref{oscillating_FCs}) and noting that $K_\mu \approx -1$ for the ground-state hyperfine transition in muonium, we estimate the sensitivity of this existing dataset to $\Lambda_\mu$ and $\Lambda'_\mu$ as shown by the solid blue lines in Fig.~\ref{Fig:Scalar_DM_Sensitivity_Plot}. 
More recent low-field measurements were performed in the MuSEUM experiment at J-PARC with a precision of $\Delta \nu / \nu \approx 9 \times 10^{-7}$ \cite{J-PARC_2021_Muonium_HFS-ZF}, where only a single line scan was performed in $\approx 15$ hours. 
The continuing MuSEUM experiment at J-PARC is targetting an improved precision of $\Delta \nu / \nu \approx 1.2 \times 10^{-9}$ with $\approx 40$ days of high-field measurements \cite{J-PARC_2021_Muonium_HFS-ZF}. 
Assuming that a single line scan takes $\approx 6$ hours \cite{Kanda_2022_Private_Comm}, we estimate the sensitivity of these ongoing measurements to $\Lambda_\mu$ and $\Lambda'_\mu$ as shown by the dashed blue lines in Fig.~\ref{Fig:Scalar_DM_Sensitivity_Plot}. 

Recent measurements of the $2S-2P$ Lamb shift in muonium were performed in the Mu-MASS experiment at PSI with a precision of $\Delta \nu / \nu \approx 2.4 \times 10^{-3}$ \cite{PSI_2022_Muonium_2S-2P}. 
Data taking took place continuously over 2 days, with a single line scan taking $\approx 3$ hours. 
The continuing Mu-MASS experiment at PSI is targetting an improved precision of $\Delta \nu / \nu \approx 10^{-5}$ with $\approx 10$ days of measurements \cite{PSI_2021_Muonium_2S-2P}. 
%Assuming the same time for a single line scan, then using Eqs.~(\ref{oscillating_FCs}) and (\ref{Mu_2S-2P}), we estimate the sensitivity of these ongoing measurements to $\Lambda_\mu$ as shown by the dashed purple line in Fig.~\ref{Fig:Scalar_DM_Sensitivity_Plot}. 
True muonium has yet to be observed directly, but there are plans to measure its ground-state hyperfine splitting interval with a precision of $\Delta \nu / \nu \sim 10^{-4}$ \cite{Crivelli_2021_TM_talk_PBC,Paolo_2022_Private_Comm}.

%%%
\emph{Hydrogen-like muonic atoms:} 
Measurements of the $2S-2P$ Lamb shift have been performed in various H-like muonic atoms, including muonic hydrogen \cite{Pohl_2010_Muonic-H,Antognini_2013_Muonic-H}, muonic deuterium \cite{Pohl_2016_Muonic-D} and muonic helium \cite{Krauth_2021_Muonic-He}, with a typical precision of $\Delta \nu / \nu \sim 10^{-5}$. 
These experiments were performed over the span of several years, with each experiment collecting $\sim 10$ days of data. 
Using Eq.~(\ref{oscillating_FCs}) and noting that $K_\mu \approx 1$ for the $2S-2P$ Lamb shift in H-like muonic atoms, we estimate the sensitivity of this combined existing dataset to $\Lambda_\mu$ and $\Lambda'_\mu$ as shown by the solid green lines in Fig.~\ref{Fig:Scalar_DM_Sensitivity_Plot}. 
Measurements of $1S-2P$ lines in H-like muonic gold have recently been performed in the muX experiment at PSI with a precision of $\Delta \nu / \nu \sim 10^{-3}$ \cite{muX_2019_Muonic-Au}. 
The CREMA collaboration aims to measure the ground-state hyperfine splitting interval in muonic hydrogen with a precision of $\Delta \nu / \nu \approx 10^{-6}$ in about 2 weeks of measurements \cite{Amaro_2022_Muonic-H_HFS, Aldo_Private_Comm}.

%%%%

%%%%%
\textbf{Indirect signatures} --- 
The exchange of a virtual scalar boson $\phi$ between SM particles generates a potential between the SM particles. 
For simplicity, let us assume the flavour-dependent hierarchy of scales $\Lambda_\mu \ll \Lambda_e \ll \Lambda_\textrm{other}$, where $\Lambda_\textrm{other}$ denotes new-physics energy scales associated with SM particles other than the electron and muon. 
The exchange of $\phi$ between a non-relativistic electron and muon generates the Yukawa-type potential $V(r) = - m_e m_\mu \exp(-m_\phi r) / (4 \pi \Lambda_e \Lambda_\mu r)$, where $r$ is the distance between the electron and muon. 
Integrating over a homogeneous spherical source body of radius $R$ and electron number density $n_e$, one finds: 
\begin{equation}
\label{scalar-mediated_potential_finite-size}
V (r) = - \frac{n_e m_e m_\mu [m_\phi R \cosh(m_\phi R) - \sinh(m_\phi R)] \exp(-m_\phi r)}{m_\phi^3 \Lambda_e \Lambda_\mu r}  \, . 
\end{equation}
In the limiting case when $m_\phi R \ll 1$, the potential in (\ref{scalar-mediated_potential_finite-size}) simplifies to the two-pointlike-particle form with all source electrons localised at the central point of the source body. 

\begin{figure*}[t!]
\centering
\includegraphics[width=8.5cm]{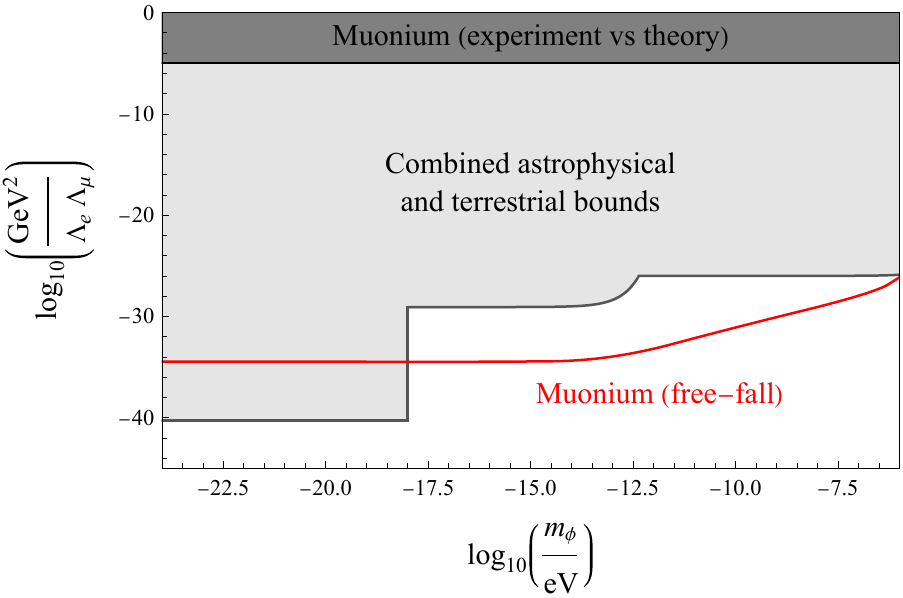}
\caption{ \normalsize (Colour online) 
Estimated sensitivities to the product of the linear scalar-electron and scalar-muon interaction parameters, $\Lambda_e \Lambda_\mu$, defined in Eq.~(\ref{linear_scalar_interactions}), of possible searches for an equivalence-principle-violating force in ongoing muonium free-fall experiments (red line). 
Complementary bounds from the comparison of the observed and predicted values of the $1S-2S$ interval in muonium are shown by the dark grey region \cite{PSI_2022_Muonium_2S-2P}, while more indirect bounds from the combination of astrophysical observations and laboratory tests of the equivalence principle are shown by the light grey region. 
Sensitivities and limits are given at the $1\sigma$ level. 
See the main text and Appendix~\ref{Appendix:C} for more details about the sensitivity estimates and complementary limits, respectively. 
}
\label{Fig:Scalar_Indirect_Sensitivity_Plot}
\end{figure*}

A test body of mass $m_\textrm{test}$ containing a single (anti)muon, such as ordinary muonium or a H-like muonic atom, would experience the additional acceleration $\boldsymbol{a}_\mu (r) = -\boldsymbol{\nabla} V(r) / m_\textrm{test}$. 
On the other hand, ordinary test bodies devoid of (anti)muons would experience much smaller additional accelerations, thereby violating the equivalence principle. 
For example, the value of the local gravitational acceleration due to Earth, $\boldsymbol{g}$, measured in free-fall experiments using muonium test atoms would differ from experiments using non-muonic test masses, with the difference being $\Delta \boldsymbol{g} \approx \boldsymbol{a}_\mu$. 
The ongoing LEMING experiment at PSI \cite{Soter_2021_Muonium-gravity} and proposed MAGE experiment \cite{MAGE_2018_Muonium-gravity} aim to measure $g$ with a precision of $\Delta g / g \sim 0.1$ using free-falling muonium test atoms in $\sim 10 - 100$ days of measurements. 
Using Eq.~(\ref{scalar-mediated_potential_finite-size}) and noting that $m_\textrm{test} \approx m_\mu$ for muonium, we estimate the sensitivity of the ongoing LEMING experiment to the combination of parameters $\Lambda_e \Lambda_\mu$ as shown by the red line in Fig.~\ref{Fig:Scalar_Indirect_Sensitivity_Plot}. 
For boson masses $m_\phi \lesssim 1/R_\oplus \approx 3 \times 10^{-14}~\textrm{eV}$, the sensitivity to $\Lambda_e \Lambda_\mu$ is approximately independent of $m_\phi$. 
For boson masses $1/R_\oplus \lesssim m_\phi \lesssim 1/h$, where $h$ is the apparatus height above Earth's surface, the sensitivity to $\Lambda_e \Lambda_\mu$ scales approximately as $\propto m_\phi^{-1}$. 
Finally, for boson masses $m_\phi \gtrsim 1/h$, the sensitivity to $\Lambda_e \Lambda_\mu$ falls off exponentially with increasing $m_\phi$. 
We assume $h \sim 1~\textrm{m}$ in Fig.~\ref{Fig:Scalar_Indirect_Sensitivity_Plot}. 
We see that ongoing free-fall experiments with muonium are capable of probing values of $\Lambda_e \Lambda_\mu$ that are up to 5 orders of magnitude more stringent than complementary laboratory and astrophysical bounds.

%%%%

%%%%%
\textbf{Discussion} --- 
From Fig.~\ref{Fig:Scalar_DM_Sensitivity_Plot}, we can see that searches for apparent oscillations of $m_\mu$ with muonium spectroscopy measurements are capable of probing scalar-muon interactions that are up to 12 orders of magnitude more stringent than astrophysical bounds, which do not necessarily assume that ultralight scalars make up any fraction of the DM and hence are independent of $\rho_\phi$. 
The peak sensitivities of older spectroscopy datasets and ongoing spectroscopy measurements to $\Lambda_\mu$ and $\Lambda'_\mu$ are comparable;  however, newer ongoing measurements taken over shorter timescales may provide better sensitivity to $\Lambda_\mu$ and $\Lambda'_\mu$ in a range of higher scalar masses than older datasets of much longer duration. 
We note that the astrophysical bounds shown in Fig.~\ref{Fig:Scalar_DM_Sensitivity_Plot}, which are derived from systems involving (proto)neutron stars, only strictly apply to interactions of muons due to the asymmetric nature of the $\mu-\bar{\mu}$ population in (proto)neutron stars. 
These astrophysical bounds are weaker in the case of interactions of anti-muons, which would increase the potential reach of spectroscopy measurements of $\bar{\mu}$-containing systems like muonium if the \textit{CPT} symmetry is violated due to $\Lambda_{\bar{\mu}} \ne \Lambda_{\mu}$ or $\Lambda'_{\bar{\mu}} \ne \Lambda'_{\mu}$. 

Our proposed spectroscopy methods to probe ultralight DM via scalar-muon couplings offer significantly more reach than the storage-ring methods discussed in Refs.~\cite{BASE_2019_Axion-g,Graham_2021_Axion-g} to probe ultralight DM via the pseudoscalar-muon coupling. 
The main reason is that scalar-type couplings are non-derivative in nature and their DM-induced signatures generally grow with decreasing $m_\phi$, see Eq.~(\ref{oscillating_FCs}). 
On the other hand, pseudoscalar-type couplings are derivative in nature and their DM-induced signatures are generally independent of $m_\phi$. 
Our proposed spectroscopy approach also offers up to 2 orders of magnitude better sensitivity than the storage-ring proposal in Ref.~\cite{Ramani_2020_Scalar-g} to probe ultralight DM via scalar-muon couplings. 
Finally, we note that the screening mechanism discussed in Ref.~\cite{Hees_2018_EP} for quadratic scalar-type interactions of ultralight DM is largely evaded in the case of the quadratic scalar-muon coupling considered in our present paper, since Earth is practically devoid of muons and anti-muons.

%%%%%
\textbf{Acknowledgements} --- 
I am grateful to Paolo Crivelli, Klaus Jungmann and Sohtaro Kanda for helpful discussions and for providing helpful information about experimental details of spectroscopy measurements with muonium. 
I am grateful to Aldo Antognini for helpful discussions and for providing helpful information about experimental details of hyperfine spectroscopy measurements with muonic hydrogen. 
I am grateful to Anna Soter for helpful discussions of free-fall experiments with muonium. 
I am grateful to Edoardo Vitagliano for helpful feedback on v1 of my preprint and for pointing out the possibility of muonic forces in compact-binary systems. 
I am grateful to Jordy de Vries, Akshay Ghalsasi and Tanmay Kumar Poddar for helpful discussions. 
This work was supported by the Australian Research Council under the Discovery Early Career Researcher Award DE210101593.

%%%%%%%%%%%%%%%%%%%%%%%%%%%%%%%%%%%%%%%%%%%%%%%%%%
%\section*{}

%%%%%%
\begin{appendix}
\section{Sensitivities of transitions in muonium and muonic atoms to variations of fundamental constants}
\label{Appendix:A}

Here we estimate the relative sensitivities of selected transitions in muonium and muonic atoms to variations of $m_\mu$ and other fundamental constants. 
For simplicity, we assume that the \textit{CPT} symmetry is conserved, which implies the equality of the muon and anti-muon masses, as well as equal interaction strength of $\phi$ with muons and anti-muons. 
We calculate the relative sensitivity coefficients $K_\mu$ at leading order, neglecting higher-order corrections that are suppressed by (additional) powers of small parameters such as $\alpha$ or $m_e / m_\mu$.

%%%
\emph{Muonium:} 
Muonium is the bound state of an electron and an anti-muon, with a reduced system mass of $m_r = m_e m_\mu / (m_e + m_\mu) \approx m_e (1 - m_e / m_\mu)$, where $m_e$ is the electron mass. 
The energy of an electronic state described by the principal quantum number $n$ is approximately given by the Rydberg formula $E_n = - m_r \alpha^2 / (2 n^2)$. 
The ground-state hyperfine splitting interval in muonium is approximately given by the Fermi formula $\Delta E_\textrm{Fermi} = 8 m_r^3 \alpha^4 / (3 m_e m_\mu)$. 
The main contribution to the $2S-2P$ Lamb shift in muonium comes from the one-loop self-energy and takes the form $\Delta E_\textrm{SE} = \mathcal{O} [ m_e (m_r / m_e)^3 \alpha^5 ]$. 
The sensitivities of the $1S-2S$ interval, ground-state hyperfine splitting interval and $2S-2P$ Lamb shift in muonium to variations of the fundamental constants are hence given by the following respective formulae: 
\begin{equation}
\label{Mu_1S-2S}
\frac{\Delta \nu_{1S-2S}}{\nu_{1S-2S}} \approx 2 \frac{\Delta \alpha}{\alpha} + \frac{\Delta m_e}{m_e} + \frac{m_e}{m_\mu} \frac{\Delta m_\mu}{m_\mu}  \, , 
% \frac{\Delta \nu_{\mu^+ e^-}^{1S-2S}}{\nu_{\mu^+ e^-}^{1S-2S}}
\end{equation}
\begin{equation}
\label{Mu_HFS}
\frac{\Delta \nu_\textrm{HFS}}{\nu_\textrm{HFS}} \approx 4 \frac{\Delta \alpha}{\alpha} + 2 \frac{\Delta m_e}{m_e} - \frac{\Delta m_\mu}{m_\mu} \, , 
% \frac{\Delta \nu_{\mu^+ e^-}^\textrm{HFS}}{\nu_{\mu^+ e^-}^\textrm{HFS}}
\end{equation}
\begin{equation}
\label{Mu_2S-2P}
\frac{\Delta \nu_{2S-2P}}{\nu_{2S-2P}} \approx 5 \frac{\Delta \alpha}{\alpha} + \frac{\Delta m_e}{m_e} + 3 \frac{m_e}{m_\mu} \frac{\Delta m_\mu}{m_\mu}  \, , 
% \frac{\Delta \nu_{\mu^+ e^-}^{2S-2P}}{\nu_{\mu^+ e^-}^{2S-2P}}
\end{equation}
where the sensitivities to changes in $m_\mu$ in Eqs.~(\ref{Mu_1S-2S}) and (\ref{Mu_2S-2P}) arise due to the mild dependence of $m_r$ on $m_\mu$.

%%%
\emph{True muonium:} 
True muonium is the bound state of a muon and an anti-muon, with $m_r = m_\mu / 2$. 
The Rydberg contribution to the $1S-2S$ interval in true muonium is analogous to that in ordinary muonium. %, but the dominant contributions to the ground-state hyperfine splitting interval and the $2S-2P$ Lamb shift are different. 
The ground-state hyperfine splitting interval in true muonium is approximately given by the Fermi-type formula $\Delta E_\textrm{Fermi} = 7 m_\mu \alpha^4 / 12$, receiving comparable contributions from an annihilation-type process and the usual exchange-type process. 
The main contribution to the $2S-2P$ Lamb shift in true muonium comes from the one-loop electronic vacuum polarisation and takes the form $\Delta E_\textrm{VP} = \mathcal{O}(m_r \alpha^3)$. 
The sensitivities of the $1S-2S$ interval, ground-state hyperfine splitting interval and $2S-2P$ Lamb shift in true muonium to variations of fundamental constants are hence given by the following respective formulae: 
\begin{equation}
\label{TM_1S-2S}
\frac{\Delta \nu_{1S-2S}}{\nu_{1S-2S}} \approx 2 \frac{\Delta \alpha}{\alpha} + \frac{\Delta m_\mu}{m_\mu}  \, , 
% \frac{\Delta \nu_{\mu^+ \mu^-}^{1S-2S}}{\nu_{\mu^+ \mu^-}^{1S-2S}} 
\end{equation}
\begin{equation}
\label{TM_HFS}
\frac{\Delta \nu_\textrm{HFS}}{\nu_\textrm{HFS}} \approx 4 \frac{\Delta \alpha}{\alpha} + \frac{\Delta m_\mu}{m_\mu} \, , 
% \frac{\Delta \nu_{\mu^+ \mu^-}^\textrm{HFS}}{\nu_{\mu^+ \mu^-}^\textrm{HFS}} 
\end{equation}
\begin{equation}
\label{TM_2S-2P}
\frac{\Delta \nu_{2S-2P}}{\nu_{2S-2P}} \approx 3 \frac{\Delta \alpha}{\alpha} + \frac{\Delta m_\mu}{m_\mu}  \, . 
% \frac{\Delta \nu_{\mu^+ \mu^-}^{2S-2P}}{\nu_{\mu^+ \mu^-}^{2S-2P}} 
\end{equation}

%%%
\emph{Hydrogen-like muonic atoms:} 
H-like muonic atoms are bound states of a single muon and a hadronic nucleus, with $m_r = m_\mu M_\textrm{nucl} / (m_\mu + M_\textrm{nucl}) \approx m_\mu$, where $M_\textrm{nucl}$ is the nuclear mass. 
The Rydberg contribution to the $1S-2S$ interval in muonic atoms is analogous to that in muonium. 
The ground-state hyperfine splitting interval in muonic atoms is similar to that in ordinary muonium and reads $\Delta E_\textrm{Fermi} = 4 Z^3 \alpha^4 m_r^3 \mu_\textrm{nucl} / (3 m_\mu m_p) \times (2I+1)/(2I)$, where $Z$ is the proton number, $\mu_\textrm{nucl}$ is the nuclear magnetic moment in units of the nuclear magneton, $m_p$ is the proton mass, and $I$ is the nuclear spin. 
Like in true muonium, the main contribution to the $2S-2P$ Lamb shift in muonic atoms comes from the one-loop electronic vacuum polarisation and takes the form $\Delta E_\textrm{VP} = \mathcal{O} (m_r Z^2 \alpha^3)$. 
Assuming that $Z \alpha \ll 1$, the sensitivities of the $1S-2S$ interval, ground-state hyperfine splitting interval and $2S-2P$ Lamb shift in H-like muonic atoms to variations of fundamental constants are hence given by the following respective formulae: 
\begin{equation}
\label{Mu_Atom_1S-2S}
\frac{\Delta \nu_{1S-2S}}{\nu_{1S-2S}} \approx 2 \frac{\Delta \alpha}{\alpha} + \frac{\Delta m_\mu}{m_\mu}  \, , 
% \frac{\Delta \nu_{\mu^+ \mu^-}^{1S-2S}}{\nu_{\mu^+ \mu^-}^{1S-2S}} 
\end{equation}
\begin{equation}
\label{Mu_Atom_HFS}
\frac{\Delta \nu_\textrm{HFS}}{\nu_\textrm{HFS}} \approx 4 \frac{\Delta \alpha}{\alpha} + 2 \frac{\Delta m_\mu}{m_\mu} - \frac{\Delta m_p}{m_p} + \frac{\Delta \mu_\textrm{nucl}}{\mu_\textrm{nucl}} \, , 
\end{equation}
\begin{equation}
\label{Mu_Atom_2S-2P}
\frac{\Delta \nu_{2S-2P}}{\nu_{2S-2P}} \approx 3 \frac{\Delta \alpha}{\alpha} + \frac{\Delta m_\mu}{m_\mu}  \, . 
% \frac{\Delta \nu_{\mu^+ \mu^-}^{2S-2P}}{\nu_{\mu^+ \mu^-}^{2S-2P}} 
\end{equation}
The $\Delta \mu_\textrm{nucl}/\mu_\textrm{nucl}$ term in Eq.~(\ref{Mu_Atom_HFS}) typically provides a small sensitivity to variations of the light-quark masses \cite{Tedesco_2006_Light-quarks}.

%\vspace{50mm}

%%%%%%
\section{Searches for signals induced by an oscillating dark-matter field}
\label{Appendix:B}

%Here we discuss some general details of searches for signals induced by an oscillating DM field. 

According to Eq.~(\ref{oscillating_FCs}), %of the main text, 
searches for DM-induced oscillations of $m_\mu$ are directly sensitive to the combination of parameters $\sqrt{\rho_\phi}/ \Lambda_\mu$ or $\rho_\phi / (\Lambda'_\mu)^2$ rather than to $\Lambda_\mu$ or $\Lambda'_\mu$. 
In order to infer information about $\Lambda_\mu$ or $\Lambda'_\mu$, one must make assumptions about the local value of $\rho_\phi$ during the course of the measurements.
In the main text, we assume the simplest possibility, namely that $\rho_\phi$ equals the average local Galactic DM density, $\rho_\phi = \rho_\textrm{DM} \approx 0.4~\textrm{GeV/cm}^3$ \cite{PDG_2022_review}. 
 In this case, $\phi_0$ (and hence $\rho_\phi$) is expected to remain approximately constant during measurements on timescales shorter than $\tau_\textrm{coh}$, implying a coherent signal of constant amplitude and fixed phase. 
However, since $\phi_0$ is expected to fluctuate stochastically on timescales greater than $\tau_\textrm{coh}$ \cite{Derevianko_2018_Stochastic-DM,Safdi_2018_Stochastic-DM}, sampling only a single value of $\phi_0$ with measurements on a short timescale may cause statistical biasing in the analysis \cite{Centers_2021_Stochastic-DM}. 
In the standard halo model, such partial sampling is only expected to degrade the sensitivity to $\Lambda_\mu$ by a factor of $\approx 1.5$ at the $1 \sigma$ level and by a similarly mild factor in the case of $\Lambda'_\mu$. 

One of the main factors determining the sensitivity of spectroscopy measurements to $\Lambda_\mu$ or $\Lambda'_\mu$ is their \textit{relative precision} rather than their (absolute) \textit{accuracy}. 
This is because one looks for \textit{changes} in transition frequencies instead of comparing the (mean) experimental value of the transition frequency with its prediction within the SM. 
The considered spectroscopy measurements are inherently broadband, with a peak sensitivity to $\Lambda_\mu$ or $\Lambda'_\mu$ typically occurring when the total duration of the measurements, $t_\textrm{dataset}$, is comparable to the signal period $T_\textrm{signal}$, with $T_\textrm{signal} = T_\textrm{osc}$ in the case of the linear-in-$\phi$ interaction and $T_\textrm{signal} = T_\textrm{osc}/2$ in the case of the quadratic-in-$\phi$ interaction. 
For a temporally coherent signal, the sensitivity scales as $\Lambda_\mu \propto m_\phi^{-1}$ and $\Lambda'_\mu \propto m_\phi^{-1}$ for $t_\textrm{cycle} \lesssim T_\textrm{signal} \lesssim t_\textrm{dataset}$, where $t_\textrm{cycle}$ is the time taken to perform a single measurement of the transition frequency, and as $\Lambda_\mu \propto m_\phi^{-2}$ and $\Lambda'_\mu \propto m_\phi^{-3/2}$ for $T_\textrm{signal} \lesssim t_\textrm{cycle}$ when the transition frequency would be affected by at most one of many largely-cancelling DM oscillations (though the scaling with $m_\phi$ may be more favourable if one also considers possible alterations of the transition lineshape due to effects of the DM field when scanning the transition line). 
When $T_\textrm{signal} \gtrsim t_\textrm{dataset}$, the sensitivity generally scales as $\Lambda_\mu \propto m_\phi$ and $\Lambda'_\mu \propto m_\phi^0$, since one cannot exclude the possibility of being near an antinode, rather than a node, of the cosinusoidal signal during the course of the measurements. 
In the temporally coherent regime, the signal-to-noise ratio (SNR) is expected to improve with the integration time $t_\textrm{int}$ as $\textrm{SNR} \propto t_\textrm{int}^{1/2}$, whereas in the temporally incoherent regime, the SNR is expected to improve as $\textrm{SNR} \propto t_\textrm{int}^{1/4} \tau_\textrm{coh}^{1/4}$. 
Hence when transitioning from the temporally coherent regime to the temporally incoherent regime, the sensitivity is expected to degrade faster with increasing $m_\phi$ by an additional factor $\propto m_\phi^{-1/4}$ in the case of $\Lambda_\mu$ and $\propto m_\phi^{-1/8}$ in the case of $\Lambda'_\mu$.

%\vspace{50mm}

%%%%%%
\section{Complementary bounds on scalar-muon interactions}
\label{Appendix:C}

%Here we present estimates of complementary bounds on scalar-muon interactions. 

%%%
\emph{Supernova cooling bounds on $\Lambda_\mu$ and $\Lambda'_\mu$:} 
The scalar-muon couplings in Eqs.~(\ref{linear_scalar_interactions}) and (\ref{quadratic_scalar_interactions}) %of the main text 
provide additional possible energy-loss channels involving the emission of $\phi$ bosons from the interiors of hot media such as proto-neutron stars in supernovae. 
In the case of the linear coupling, the consideration of the emission of $\phi$ bosons from the interior of supernova SN1987a rules out the range of parameters $10^3~\textrm{GeV} \lesssim \Lambda_\mu \lesssim 2 \times 10^7~\textrm{GeV}$ for $m_\phi \lesssim T$, where $T \sim 30~\textrm{MeV}$ is the peak core temperature of the proto-neutron star \cite{Caputo_2022_SNe-muon_bounds}. 
Noting that the muons in a proto-neutron star are semi-degenerate and that the typical energy of an emitted $\phi$ boson is $\sim T$, we can estimate the ratio of the energy-loss rates due to the semi-Compton-type processes $\mu+p \to \mu+p+2\phi$ for the quadratic coupling and $\mu+p \to \mu+p+\phi$ for the linear coupling as follows: 
\begin{equation}
\label{SNe_ratio_relation}
\frac{\epsilon^\textrm{quad}_{2\phi}}{\epsilon^\textrm{lin}_{\phi}} \sim \frac{T^2 / (\Lambda'_\mu)^4}{1 / \Lambda_\mu^2} \, . 
\end{equation}
%Noting that the consideration of the emission of $\phi$ bosons from the interior of supernova SN1987a rules out the range of parameters $10^3~\textrm{GeV} \lesssim \Lambda_\mu \lesssim 2 \times 10^7~\textrm{GeV}$ for $m_\phi \lesssim 30~\textrm{MeV}$ in the case of the linear coupling \cite{Caputo_2022_SNe-muon_bounds}, 
We hence estimate that the following rough range of parameters is ruled out in the case of the quadratic coupling: 
\begin{equation}
\label{SNe_quad_bound}
10~\textrm{GeV} \lesssim \Lambda'_\mu \lesssim 10^3~\textrm{GeV} \, , ~\textrm{for}~ m_\phi \lesssim 15~\textrm{MeV} \, . 
\end{equation}

%%%
\emph{Compact binary system bounds on $\Lambda_\mu$ and $\Lambda'_\mu$:} 
The scalar-muon couplings in Eqs.~(\ref{linear_scalar_interactions}) and (\ref{quadratic_scalar_interactions}) %of the main text 
provide additional possible channels for the decay of the orbital period of a compact binary system containing at least one neutron star via the emission of $\phi$-boson radiation. 
In the case of the linear coupling, the radiation of $\phi$ bosons from the muon content of a neutron star in a compact binary system would affect the decay rate of the orbital period of the system if $m_\phi < n \Omega$, where $\Omega$ is the orbital frequency of the binary system, with $n=1$ corresponding to the fundamental mode and $n=2,3,...$ corresponding to higher-order modes that contribute in the case of non-circular orbits. 
Refs.~\cite{Poddar_2019_binary,Laha_2020_binary} derived bounds on vector-type couplings to muons at the level $g_\mu^V \lesssim 10^{-20} - 10^{-19}$ for boson masses $\lesssim 10^{-18}~\textrm{eV}$. 
One expects comparable limits on the linear scalar-muon coupling, $g_\mu^V \sim g_\mu^s \equiv m_\mu / \Lambda_\mu$, implying: 
\begin{equation}
\label{binary_lin_bound}
\Lambda_\mu \gtrsim 10^{18} - 10^{19}~\textrm{GeV} \, , ~\textrm{for}~ m_\phi \lesssim 10^{-18}~\textrm{eV} \, . 
\end{equation}

The emission of a pair of $\phi$ bosons is possible when $2 m_\phi < n \Omega$. %, where $\Omega$ is the orbital frequency of the binary system, with $n=1$ corresponding to the fundamental mode and $n=2,3,...$ corresponding to higher-order modes. 
In the limit when $\Omega a \ll 1$ (where $a$ denotes the semi-major axis of the binary system), the dominant radiation mechanism is dipole type as long as the muon-to-neutron number ratio is not too similar in the two bodies. 
Noting that the typical energy of a radiated $\phi$ boson is $\sim \Omega$, we can estimate the ratio of power loss rates due to the emission of a pair of $\phi$ bosons in the case of the quadratic coupling and of a single $\phi$ boson in the case of the linear coupling via the following Larmor-type estimate: 
\begin{equation}
\label{binary_ratio_relation}
\frac{P^\textrm{quad}_{2\phi}}{P^\textrm{lin}_{\phi}} \sim \frac{\Omega^2 / (\Lambda'_\mu)^4}{1 / \Lambda_\mu^2} \, . 
\end{equation}
We hence arrive at the following order-of-magnitude estimate for the bound on the quadratic coupling: 
\begin{equation}
\label{binary_quad_bound}
\Lambda'_\mu \gtrsim 10^{-4}~\textrm{GeV} \, , ~\textrm{for}~ m_\phi \lesssim 0.5 \times 10^{-18}~\textrm{eV} \, . 
\end{equation}
There is a similarly mild bound on $\Lambda'_\mu$ from the consideration of the effect on the orbital dynamics of the binary neutron-star system due to the exchange of a pair of $\phi$ bosons between the muon contents of the two neutron stars, which induces the modified Yukawa-type potential $V_{2\phi}(r) \approx -m_\mu^2 / [64 \pi^3 (\Lambda'_\mu)^4 r^3]$ in the limit when $2 m_\phi r \ll 1$.

%%%
\emph{Bounds on the combination of parameters $\Lambda_e \Lambda_\mu$:} 
Consideration of intra-atomic exchange of a virtual scalar boson $\phi$ on the $1S-2S$ interval in muonium places the bound $\Lambda_e \Lambda_\mu \gtrsim 10^5~\textrm{GeV}^2$ for $m_\phi \ll m_e \alpha$ \cite{PSI_2022_Muonium_2S-2P}, as shown in Fig.~\ref{Fig:Scalar_Indirect_Sensitivity_Plot} %of the main text 
by the dark grey region.\footnote{\normalsize Similar approaches using muonium spectroscopy data have also been used to constrain pseudoscalar \cite{Frugiuele_2019_Muonium_pseudoscalar} and vector \cite{Karshenboim_2014_Muonium_vector} couplings.} 
Feebler interactions can be indirectly constrained by combining the astrophysical bounds on $\Lambda_\mu$ discussed above with laboratory bounds on $\Lambda_e$. 
Searches for $\phi$-mediated equivalence-principle-violating forces place the limit $\Lambda_e \gtrsim 6 \times 10^{21}~\textrm{GeV}$ for $m_\phi \lesssim 3 \times 10^{-14}~\textrm{eV}$ based on data from the space-based MICROSCOPE mission \cite{MICROSCOPE_2022_EP} assuming that the elemental composition of Earth is a 1:1:1 ratio of $^{24}$Mg$^{16}$O, $^{28}$Si$^{16}$O$_{2}$ and $^{56}$Fe by number and $\Lambda_e \gtrsim 5 \times 10^{18}~\textrm{GeV}$ for $m_\phi \lesssim 2 \times 10^{-6}~\textrm{eV}$ based on data from the ground-based Rot-Wash experiment \cite{Rot-Wash_1999_EP,Leefer_2016_EP}. 
The combination of these astrophysical and laboratory bounds is shown in Fig.~\ref{Fig:Scalar_Indirect_Sensitivity_Plot} %of the main text 
by the light grey region.

%%%
\emph{Indirect bounds on $\Lambda_\mu$ from equivalence-principle tests:} 
The scalar-lepton coupling in Eq.~(\ref{linear_scalar_interactions}) %of the main text 
radiatively generates a scalar-diphoton coupling via a triangular lepton loop, with an effective scalar-diphoton coupling of the size $1/\Lambda_\gamma \sim (\alpha / \pi) (1/ \Lambda_l) [1+\mathcal{O}(m_\phi^2 / m_l^2)]$ when $m_\phi \ll m_l$, with the coefficients being independent of the lepton species $l$. 
This suggests the peak figure-of-merit sensitivities (at $1\sigma$ level) of $\Lambda_\mu \sim 3 \times 10^{20}~\textrm{GeV}$ for $m_\phi \lesssim 3 \times 10^{-14}~\textrm{eV}$ based on the MICROSCOPE data in Ref.~\cite{MICROSCOPE_2022_EP} and $\Lambda_\mu \sim 4 \times 10^{17}~\textrm{GeV}$ for $m_\phi \lesssim 2 \times 10^{-6}~\textrm{eV}$ based on the Rot-Wash data in Ref.~\cite{Rot-Wash_1999_EP}. 
However, these indirect bounds on $\Lambda_\mu$ from torsion-pendula experiments can be practically evaded, e.g., due to an analogous scalar-tau coupling with $\Lambda_\tau = \Lambda_\mu$ (or $\Lambda_\tau \approx \Lambda_\mu$), but of opposite sign to the scalar-muon coupling, when $m_\phi \ll m_\mu$; 
on the other hand, direct probes of the scalar-muon and scalar-tau couplings would remain sensitive in this case. 

\end{appendix}

%===================================================================================


\begin{thebibliography}{99}
\normalsize

\bibitem{PDG_2022_review} R.~L.~Workman \textit{et al}.~(Particle Data Group), \textit{The Review of Particle Physics}, Prog.~Theor.~Exp.~Phys.~\textbf{2022}, 083C01 (2022). 

\bibitem{Roszkowski_2018_WIMP-review} L.~Roszkowski, E.~M.~Sessolo, and S.~Trojanowski, \textit{WIMP dark matter candidates and searches --- current status and future prospects}, Rep.~Prog.~Phys.~\textbf{81}, 066201 (2018). 

\bibitem{Kim-Carosi_2010_Axions-review} J.~E.~Kim and G.~Carosi, \textit{Axions and the strong CP problem}, Rev.~Mod.~Phys.~\textbf{82}, 557 (2010). 

\bibitem{Graham_2015_Relaxion} P.~W.~Graham, D.~E.~Kaplan, and S.~Rajendran, \textit{Cosmological Relaxation of the Electroweak Scale}, Phys.~Rev.~Lett.~\textbf{115}, 221801 (2015). 

\bibitem{Snowmass_2022_Scalars+Vectors} D. Antypas \textit{et al}., \textit{New Horizons:~Scalar and Vector Ultralight Dark Matter}, arXiv:2203.14915. 

\bibitem{Rot-Wash_1999_EP} G.~L.~Smith, C.~D.~Hoyle, J.~H.~Gundlach, E.~G.~Adelberger, B.~R.~Heckel, and H.~E.~Swanson, \textit{Short-range tests of the equivalence principle}, Phys.~Rev.~D \textbf{61}, 022001 (1999). 

\bibitem{Eot-Wash_2008_EP} S.~Schlamminger, K.-Y.~Choi, T.~A.~Wagner, J.~H.~Gundlach, and E.~G.~Adelberger, \textit{Test of the Equivalence Principle Using a Rotating Torsion Balance}, Phys.~Rev.~Lett.~\textbf{100}, 041101 (2008). 

\bibitem{Wagner_2012_EP-review} T.~A.~Wagner, S.~Schlamminger, J.~H.~Gundlach, and E.~G.~Adelberger, \textit{Torsion-balance tests of the weak equivalence principle}, Class.~Quantum Grav.~\textbf{29}, 184002 (2012). 

\bibitem{Leefer_2016_EP} N.~Leefer, A.~Gerhardus, D.~Budker, V.~V.~Flambaum, and Y.~V.~Stadnik, \textit{Search for the Effect of Massive Bodies on Atomic Spectra and Constraints on Yukawa-Type Interactions of Scalar Particles}, Phys.~Rev.~Lett.~\textbf{117}, 271601 (2016). 

\bibitem{MICROSCOPE_2017_EP} P.~Touboul \textit{et al}., \textit{MICROSCOPE Mission:~First Results of a Space Test of the Equivalence Principle}, Phys.~Rev.~Lett.~\textbf{119}, 231101 (2017). 

\bibitem{Hees_2018_EP} A.~Hees, O.~Minazzoli, E.~Savalle, Y.~V.~Stadnik, and P.~Wolf, \textit{Violation of the equivalence principle from light scalar dark matter}, Phys.~Rev.~D~\textbf{98}, 064051 (2018). 

\bibitem{MICROSCOPE_2022_EP} P.~Touboul \textit{et al}., \textit{MICROSCOPE Mission:~Final Results of the Test of the Equivalence Principle}, Phys.~Rev.~Lett.~\textbf{129}, 121102 (2022). 

\bibitem{Pohl_2010_Muonic-H} R.~Pohl \textit{et al}., \textit{The size of the proton}, Nature~\textbf{466}, 213 (2010). 

\bibitem{Antognini_2013_Muonic-H} A.~Antognini \textit{et al}., \textit{Proton Structure from the Measurement of 2S-2P Transition Frequencies of Muonic Hydrogen}, Science~\textbf{339}, 417 (2013). 

\bibitem{Pohl_2016_Muonic-D} R.~Pohl \textit{et al}.~(The CREMA Collaboration), \textit{Laser spectroscopy of muonic deuterium}, Science~\textbf{353}, 669 (2016). 

\bibitem{Krauth_2021_Muonic-He} J.~J.~Krauth \textit{et al}., \textit{Measuring the $\alpha$-particle charge radius with muonic helium-4 ions}, Nature~\textbf{589}, 527 (2021). 

\bibitem{Pospelov_2011_Proton_radius} B.~Batell, D.~McKeen, and M.~Pospelov, \textit{New Parity-Violating Muonic Forces and the Proton Charge Radius}, Phys.~Rev.~Lett.~\textbf{107}, 011803 (2011). 

\bibitem{BNL_2006_Muon_g-2} G.~W.~Bennett \textit{et al}.~(Muon g-2 Collaboration), \textit{Final report of the E821 muon anomalous magnetic moment measurement at BNL}, Phys.~Rev.~D \textbf{73}, 072003 (2006). 

\bibitem{FNAL_2021_Muon_g-2} B.~Abi \textit{et al.}~(Muon g-2 Collaboration), \textit{Measurement of the Positive Muon Anomalous Magnetic Moment to 0.46 ppm}, Phys.~Rev.~Lett.~\textbf{126}, 141801 (2021). 

\bibitem{Pospelov_2017_Muon_g-2} C.-Y.~Chen, M.~Pospelov, and Y.-M.~Zhong, \textit{Muon beam experiments to probe the dark sector}, Phys.~Rev.~D \textbf{95}, 115005 (2017). 

\bibitem{LAMPF_1975_Muonium_HFS-ZF} D.~E.~Casperson \textit{et al}., \textit{A new high precision measurement of the muonium hyperfine structure interval $\Delta \nu$}, Phys.~Lett.~B \textbf{59}, 397 (1975). 

\bibitem{LAMPF_1999_Muonium_HFS-HF} W.~Liu \textit{et al}., \textit{High Precision Measurements of the Ground State Hyperfine Structure Interval of Muonium and of the Muon Magnetic Moment}, Phys.~Rev.~Lett.~\textbf{82}, 711 (1999). 

\bibitem{RAL_2000_Muonium_1S-2S} V.~Meyer \textit{et al}., \textit{Measurement of the 1s-2s Energy Interval in Muonium}, Phys.~Rev.~Lett.~\textbf{84}, 1136 (2000). 

\bibitem{J-PARC_2021_Muonium_HFS-ZF} S.~Kanda \textit{et al}., \textit{New precise spectroscopy of the hyperfine structure in muonium with a high-intensity pulsed muon beam}, Phys.~Lett.~B \textbf{815}, 136154 (2021). 

\bibitem{PSI_2022_Muonium_2S-2P} B.~Ohayon \textit{et al}.~(Mu-MASS Collaboration), \textit{Precision Measurement of the Lamb Shift in Muonium}, Phys.~Rev.~Lett.~\textbf{128}, 011802 (2022). 

\bibitem{Wilczek_1983_Axion-cosmo} J.~Preskill, M.~B.~Wise, and F.~Wilczek, \textit{Cosmology of the invisible axion}, Phys.~Lett.~B~\textbf{120}, 127 (1983). 

\bibitem{Sikivie_1983_Axion-cosmo} L.~F.~Abbott and P.~Sikivie, \textit{A cosmological bound on the invisible axion}, Phys.~Lett.~B~\textbf{120}, 133 (1983). 

\bibitem{Fischler_1983_Axion-cosmo} M.~Dine and W.~Fischler, \textit{The not-so-harmless axion}, Phys.~Lett.~B~\textbf{120}, 137 (1983). 

\bibitem{Batell_2021_Thermal_misalignment} B.~Batell and A.~Ghalsasi, arXiv:2109.04476. 

\bibitem{Piazza_2020_scalar-Higgs} F.~Piazza and M.~Pospelov, \textit{Sub-eV scalar dark matter through the super-renormalizable Higgs portal}, Phys.~Rev.~D~\textbf{82}, 043533 (2010). 

\bibitem{Moody_1984_scalar_CPV} J.~E.~Moody and F.~Wilczek, \textit{New macroscopic forces?}, Phys.~Rev.~D~\textbf{30}, 130 (1984). 

\bibitem{deVries_2022_scalar_CPV} W.~Dekens, J.~de Vries and S.~Shain, \textit{CP-violating axion interactions in effective field theory}, J.~High Energy Phys.~\textbf{2022}, 14 (2022). 

\bibitem{Stadnik_2015_DM-LIFO} Y.~V.~Stadnik and V.~V.~Flambaum, \textit{Searching for Dark Matter and Variation of Fundamental Constants with Laser and Maser Interferometry}, Phys.~Rev.~Lett.~\textbf{114}, 161301 (2015). 

\bibitem{Stadnik_2015_DM-VFCs} Y.~V.~Stadnik and V.~V.~Flambaum, \textit{Can Dark Matter Induce Cosmological Evolution of the Fundamental Constants of Nature?}, Phys.~Rev.~Lett.~\textbf{115}, 201301 (2015). 

\bibitem{RAL_2000_Thesis-Meyer} V.~Meyer, \textit{Measurement of the Muon Mass by Doppler-Free Two-Photon Spectroscopy of the 1s-2s Transition in Muonium, Hydrogen and Deuterium}, PhD Thesis, Heidelberg, 1998. 

\bibitem{PSI_2018_Muonium_1S-2S} P.~Crivelli, \textit{The Mu-MASS (muonium laser spectroscopy) experiment}, Hyperfine Interact.~\textbf{239}, 49 (2018). 

\bibitem{Paolo_2022_Private_Comm} P.~Crivelli, \textit{Private Communication}. 

\bibitem{LAMPF_1999_Thesis-Liu} W.~Liu, \textit{High precision measurement of muonium ground state hyperfine interval and muon magnetic moment}, PhD Thesis, Yale, 1997. 

\bibitem{Kanda_2022_Private_Comm} S.~Kanda, \textit{Private Communication}. 

\bibitem{LAMFP_1999_Thesis-Reinhard} I.~Reinhard, \textit{Precision Spectroscopy of the Muonium Hyperfine Structure with the ``Old Muonium'' Technique}, PhD Thesis, Heidelberg, 1996. 

\bibitem{PSI_2021_Muonium_2S-2P} G.~Janka, B.~Ohayon, and P.~Crivelli, \textit{Muonium Lamb shift:~theory update and experimental prospects}, EPJ Web Conf.~\textbf{262}, 01001 (2022). 

\bibitem{Crivelli_2021_TM_talk_PBC} P.~Crivelli, \textit{Progress on the true muonium front}, talk at the ``PBC General Meeting'', CERN, $3^\textrm{rd}$ Dec 2021. 

\bibitem{muX_2019_Muonic-Au} A.~Skawran, \textit{Towards nuclear structure with radioactive muonic atoms}, Il Nuovo Cimento~\textbf{42 C} (2019) 125. 

\bibitem{Amaro_2022_Muonic-H_HFS} P.~Amaro \textit{et al}., \textit{Laser excitation of the 1s-hyperfine transition in muonic hydrogen}, SciPost Phys.~\textbf{13}, 020 (2022). 

\bibitem{Aldo_Private_Comm} A.~Antognini, \textit{Private Communication}. 

\bibitem{Viel_2017_Fuzzy-DM} V.~Irsic, M.~Viel, M.~G.~Haehnelt, J.~S.~Bolton, and G.~D.~Becker, \textit{First Constraints on Fuzzy Dark Matter from Lyman-$\alpha$ Forest Data and Hydrodynamical Simulations}, Phys.~Rev.~Lett.~\textbf{119}, 031302 (2017). 

\bibitem{Viel_2019_Fuzzy-DM} M.~Nori, R.~Murgia, V.~Irsic, M.~Baldi, and M.~Viel, \textit{Lyman $\alpha$ forest and non-linear structure characterization in Fuzzy Dark Matter cosmologies}, Mon.~Notices Royal Astron.~Soc.~\textbf{482}, 3227 (2019). 

\bibitem{Marsh_2019_Fuzzy-DM} D.~J.~E.~Marsh and J.~C.~Niemeyer, \textit{Strong Constraints on Fuzzy Dark Matter from Ultrafaint Dwarf Galaxy Eridanus II}, Phys.~Rev.~Lett.~\textbf{123}, 051103 (2019). 

\bibitem{Schutz_2020_Fuzzy-DM} K.~Schutz, \textit{Subhalo mass function and ultralight bosonic dark matter}, Phys.~Rev.~D~\textbf{101}, 123026 (2020). 

\bibitem{Soter_2021_Muonium-gravity} A.~Soter and A.~Knecht, \textit{Development of a cold atomic muonium beam for next generation atomic physics and gravity experiments}, SciPost Phys.~Proc.~\textbf{5}, 031 (2021). 

\bibitem{MAGE_2018_Muonium-gravity} A.~Antognini \textit{et al}., \textit{Studying Antimatter Gravity with Muonium}, Atoms~\textbf{6}, 17 (2018). 

\bibitem{BASE_2019_Axion-g} C.~Smorra \textit{et al}., \textit{Direct limits on the interaction of antiprotons with axion-like dark matter}, Nature~\textbf{575}, 310 (2019). 

\bibitem{Graham_2021_Axion-g} P.~W.~Graham, S.~Haciomeroglu, D.~E.~Kaplan, Z.~Omarov, S.~Rajendran, and Y.~K.~Semertzidis, \textit{Storage ring probes of dark matter and dark energy}, Phys.~Rev.~D~\textbf{103}, 055010 (2021). 

\bibitem{Ramani_2020_Scalar-g} R.~Janish and H.~Ramani, \textit{Muon g-2 and EDM experiments as muonic dark matter detectors}, Phys.~Rev.~D~\textbf{102}, 115018 (2020). 

%\bibitem{Tedesco_2006_Light-quarks} V.~V.~Flambaum and A.~F.~Tedesco, \textit{Dependence of nuclear magnetic moments on quark masses and limits on temporal variation of fundamental constants from atomic clock experiments}, Phys.~Rev.~C~\textbf{73}, 055501 (2006). 



\bibitem{Tedesco_2006_Light-quarks} V.~V.~Flambaum and A.~F.~Tedesco, \textit{Dependence of nuclear magnetic moments on quark masses and limits on temporal variation of fundamental constants from atomic clock experiments}, Phys.~Rev.~C~\textbf{73}, 055501 (2006). 

%\bibitem{PDG_2022_review} R.~L.~Workman \textit{et al}.~(Particle Data Group), \textit{The Review of Particle Physics}, Prog.~Theor.~Exp.~Phys.~\textbf{2022}, 083C01 (2022). 

\bibitem{Derevianko_2018_Stochastic-DM} A.~Derevianko, \textit{Detecting dark-matter waves with a network of precision-measurement tools}, Phys.~Rev.~A~\textbf{97}, 042506 (2018). 

\bibitem{Safdi_2018_Stochastic-DM} J.~W.~Foster, N.~L.~Rodd, and B.~R.~Safdi, \textit{Revealing the dark matter halo with axion direct detection}, Phys.~Rev.~D~\textbf{97}, 123006 (2018). 

\bibitem{Centers_2021_Stochastic-DM} G.~P.~Centers \textit{et al}., \textit{Stochastic fluctuations of bosonic dark matter}, Nat.~Commun.~\textbf{12}, 1 (2021). 

\bibitem{Caputo_2022_SNe-muon_bounds} A.~Caputo, G.~Raffelt, and E.~Vitagliano, \textit{Muonic boson limits:~Supernova redux}, Phys.~Rev.~D~\textbf{105}, 035022 (2022). 

\bibitem{Poddar_2019_binary} T.~Kumar Poddar, S.~Mohanty, and S.~Jana, \textit{Vector gauge boson radiation from compact binary systems in a gauged $L_\mu - L_\tau$ scenario}, Phys.~Rev.~D~\textbf{100}, 123023 (2019). 

\bibitem{Laha_2020_binary} J.~ A.~Dror, R.~Laha, and T.~Opferkuch, \textit{Probing muonic forces with neutron star binaries}, Phys.~Rev.~D~\textbf{102}, 023005 (2020). 

%\bibitem{PSI_2022_Muonium_2S-2P} B.~Ohayon \textit{et al}.~(Mu-MASS Collaboration), \textit{Precision Measurement of the Lamb Shift in Muonium}, Phys.~Rev.~Lett.~\textbf{128}, 011802 (2022). 

\bibitem{Frugiuele_2019_Muonium_pseudoscalar} C.~Frugiuele, J.~Perez-Rios, and C.~Peset, \textit{Current and future perspectives of positronium and muonium spectroscopy as dark sectors probe}, Phys.~Rev.~D~\textbf{100}, 015010 (2019). 

\bibitem{Karshenboim_2014_Muonium_vector} S.~G.~Karshenboim, D.~McKeen, and M.~Pospelov, \textit{Constraints on muon-specific dark forces}, Phys.~Rev.~D~\textbf{90}, 073004 (2014). 

%\bibitem{MICROSCOPE_2022_EP} P.~Touboul \textit{et al}., \textit{MICROSCOPE Mission:~Final Results of the Test of the Equivalence Principle}, Phys.~Rev.~Lett.~\textbf{129}, 121102 (2022). 

%\bibitem{Rot-Wash_1999_EP} G.~L.~Smith, C.~D.~Hoyle, J.~H.~Gundlach, E.~G.~Adelberger, B.~R.~Heckel, and H.~E.~Swanson, \textit{Short-range tests of the equivalence principle}, Phys.~Rev.~D \textbf{61}, 022001 (1999). 

%\bibitem{Leefer_2016_EP} N.~Leefer, A.~Gerhardus, D.~Budker, V.~V.~Flambaum, and Y.~V.~Stadnik, \textit{Search for the Effect of Massive Bodies on Atomic Spectra and Constraints on Yukawa-Type Interactions of Scalar Particles}, Phys.~Rev.~Lett.~\textbf{117}, 271601 (2016). 







\end{thebibliography}
\end{document}